\documentclass[10pt]{iopart}

\usepackage{setstack,mathrsfs,lineno,amssymb,amsbsy}
\usepackage{graphicx}
\usepackage{float}
\usepackage{tabularx}
\usepackage{adjustbox}
\usepackage{lipsum}
\usepackage{bbm}
\usepackage[dvipsnames]{xcolor,colortbl}
\usepackage[font=small,labelfont=bf,labelsep=period]{caption}
\usepackage[square,numbers,sort&compress]{natbib}

\usepackage[symbol]{footmisc}

\renewcommand\harvardyearright[1]{.}

\usepackage{iopams}

\expandafter\let\csname equation*\endcsname\relax

\expandafter\let\csname endequation*\endcsname\relax
\usepackage{soul}

\begin{document}

\title[Piezoelectric-based electroelastic metasurface]{Electroelastic metasurface with resonant piezoelectric shunts for tunable wavefront control}
\author{Z. Lin and S. Tol\footnote[0*]{$^{*}$Authors to whom any correspondence should be addressed.}}
\address{Department of Mechanical Engineering, University of Michigan, Ann-Arbor, MI 48109-2125, United States}
\ead{stol@umich.edu}
\date{September 2022}
\vspace{10pt}

\begin{abstract}
In this paper, we design a tunable phase-modulated metasurface composed of periodically distributed piezoelectric patches with resonant-type shunt circuits. The electroelastic metasurface can control the wavefront of the lowest antisymmetric mode Lamb wave ($A_0$ mode) in a small footprint due to its subwavelength features.~The fully coupled electromechanical model is established to study the transmission characteristics of the metasurface unit and validated through numerical and experimental studies. Based on the analysis of the metasurface unit, we first explore the performance of electroelastic metasurface with single-resonant shunts and then extend its capability with multi-resonant shunts.~By only tuning the electric loads in the shunt circuits, we utilize the proposed metasurface to accomplish wave deflection and wave focusing of $A_0$ mode Lamb waves at different angles and focal points, respectively. Numerical simulations show that the metasurface with single-resonant shunts can deflect the wavefront of 5 kHz and 6 kHz flexural waves by desired angles with less than $2\%$ deviation. In addition, it can be tuned to achieve nearly three times displacement amplification at the designed focal point for a wide range of angles from $-75^\circ$ to $75^\circ$. Furthermore, with multi-resonant shunts, the piezoelectric-based metasurface can accomplish anomalous wave control over flexural waves at multiple frequencies (i.e., simultaneously at 5 kHz and 10 kHz), developing new potentials toward a broad range of engineering applications such as demultiplexing various frequency components or guiding and focusing them at different positions.

\vspace{2pc}
\noindent{\it Keywords}: electroelastic metasurface, piezoelectric shunts, local resonance, elastic wavefront control 
\end{abstract}
\ioptwocol
\maketitle

\section{Introduction}
Piezoelectricity offers means to enable tunable control of the acoustic/elastic waves in artificially engineered periodic structures. The ease of modifying structural dynamic properties via piezoelectric materials and their suitability for applications at different geometric scales make them excellent candidates for adaptive system designs~\cite{anton2007review,chen2014metamaterials,tol2017phononic,chen2018review,ji2021recent}. Piezoelectric elements embedded in the metamaterials have been demonstrated to change the substrate’s dynamics properties by tailoring the electrical impedance in shunt circuits without any need for structural modification~\cite{hagood1991damping,thorp2001attenuation,airoldi2011design,hwan2011active}.  
To this end, researchers adopted different electric loads in the shunt circuit to tune the dispersion properties or enrich the performance of piezoelectric-based metamaterials. Inductive and resistive loads in the piezoelectric-shunt circuit were first proposed by Hagood and von Flotow~\cite{hagood1991damping} to enhance the vibration suppression performance of the piezoelectric metastructure.~Later, other piezoelectric shunts with inductance and negative capacitance were explored to design electromechanical metamaterials (MMs) and phononic crystals (PCs), which exhibited unconventional behaviors such as tunable bandgap and dispersive properties~\cite{thorp2001attenuation}, leading to diverse applications including sound/vibration mitigation~\cite{wang2010low,chen2013wave,bergamini2015hybrid,cardella2016manipulating}, energy harvesting~\cite{de2020graded,sugino2018analysis}, wave directivity manipulators~\cite{casadei2012piezoelectric,celli2015tunable,ouisse2016piezo,xu2017tunable,li2019active}, adaptive GRIN lens~\cite{xu2017adaptive,yi2016flexural} and multifunctional designs~\cite{hu2018internally,sugino2018analysis,alshaqaq2020graded,lin2021piezoelectric}. 

Emerged as a new kind of phase-modulated structure in the last decade, metasurfaces offer compact and lightweight designs, which are especially advantageous for wavefront control in the low-frequency regime~\cite{cummer2016controlling,assouar2018acoustic}. A metasurface consists of an array of subwavelength-scaled phase modulators that introduce abrupt phase shifts in the wave propagation path and tailor wavefront based on generalized Snell’s law.~While acoustic/elastic wave manipulation via metasurfaces has been explored through various design concepts~\cite{cummer2016controlling,assouar2018acoustic,su2016focusing,shen2018elastic,liu2017source,cao2018deflecting,cao2020disordered,lee2018mass,xie2014wavefront,zhu2016anomalous,lin2021elastic}, once fabricated, most of them only operate in a narrow frequency band.~This limitation can be overcome by employing piezoelectric materials in the metasurface design achieving tunable wavefront control.~For instance, real-time active control of elastic waves was demonstrated by  embedding a feedback controller in the circuit and using sensing-and-actuating units in the metasurface~\cite{chen2018programmable,chen2020active,li2021active}. In addition, analog circuit~\cite{li2018tunable,xia2019situ,yaw2021stiffness,shi2021tunable,yaw2022anomalous} were used in piezoelectric-based adaptive metasurface design without resorting to the digital controller.~To this end, Li et al.~\cite{li2018tunable} proposed an adaptive metasurface designed by five piezoelectric patches shunted with a negative capacitance to control the out-of-plane displacement field.~Later,~Xia et al.~\cite{xia2019situ} explored the capability of piezoelectric-based metasurface in manipulating the in-plane vibration modes with negative capacitance shunts.~Nonetheless, most of them rely on negative capacitance shunts, which requires synthetic electric components and additional source to power the operational amplifier in the shunt circuits and can be unstable in engineering practice~\cite{de2008vibration,chen2011band}. Recently, researchers exploited the resonant shunt consisting of inductive loads in the electroelastic metasurface design~\cite{shi2021tunable}.~Compared to the active control technique and negative capacitance shunts, these resonant-type metasurfaces have the advantage of always being stable and easy to implement, offering a reliable avenue in real-life applications~\cite{thomas2011performance}. Yet, the phase modulation capability of resonant-type electroelastic metasurfaces is less explored.

In this paper, we study piezoelectric-based metasurfaces that can be tuned by tailoring the electric impedance in the circuit via inductive loads.~We advance the state of the art by introducing multi-resonant shunts in the electroelastic metasurface design to achieve simultaneous wavefront control over multiple frequencies.~We develop a fully coupled electromechanical model of the electroelastic metasurface and validate its accuracy through numerical and experimental studies. Then, single-resonant and multi-resonant shunts are utilized in the proposed metasurface to accomplish anomalous wave control over flexural waves at desired frequencies. 

\section{Wave propagation characteristics of the metasurface unit}
\begin{figure}
\includegraphics[width=7cm]{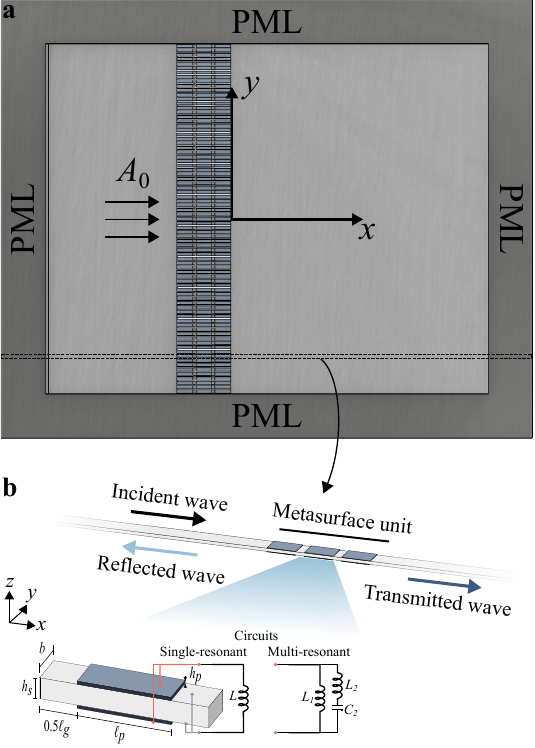}
\caption{Schematic of (a) piezoelectric-based electroelastic metasurface embedded in a plate and (b) the metasurface unit. The perfectly matched layers (PMLs) are implemented to simulate the infinite domains. The inset in (b) shows the piezoelectric bimorph with resonant shunt circuits.}
\label{fig:shcema3}
\end{figure}

We study tunable wavefront control of low-frequency elastic waves via a piezoelectric metasurface with different electric loads. The proposed piezoelectric-based electroelastic metasurface consists of 79 slender beams with a 2~mm gap and periodically placed piezoelectric patches, which are embedded in an infinite aluminum plate and used to control the $A_0$ mode Lamb wave as shown in figure~\ref{fig:shcema3}(a). 

The wave characteristics of the metasurface unit are resolved by considering an infinitely long aluminum waveguide with a beam width of $b$=5 mm and a thickness $h_s$=5 mm as shown in figure~\ref{fig:shcema3}(b). Three pairs of piezoelectric patches (PZT-5H) with a length of $\ell_p$=25 mm and a thickness of $h_p$=0.3 mm are periodically distributed on the substrate beam with a gap distance of $\ell_g$=5 mm. ~Each pair is composed of piezoelectric bimorphs (i.e., two piezoelectric patches sandwiching the central substrate beam) in parallel connection and individually shunted to an identical electric circuit. The metasurface unit has a total length of 85 mm. Therefore, it controls the flexural wave at the interested frequency of 5 kHz ($\lambda=96.1$ mm) within a subwavelength pattern.

Since the substrate beam has a relatively small thickness compared to the wavelength (i.e., $\lambda/h_s>6$), the equation of motion of the flexural wave is governed by Euler-Bernoulli theory~\cite{gonccalves2007numerical,fahy2007sound}. 

The piezoelectric patches produce a bending moment in the composite section due to the inverse piezoelectric effect~\cite{erturk2011piezoelectric}.~Coupled bending moment equation can be written as: 
\begin{equation}
\hskip1.5cm
\label{coulped moment}
M_i(x,t)=-YI\frac{\partial^2 w_i(x,t)}{\partial x^2}+\chi v_i(t)
\end{equation}
where~$\it{v_i}$~represents
the voltage across the $i^{th}$ piezoelectric bimorph.~$\chi=be_{31}(h_{p}+h_{s})$ represents the backward coupling term.~$e_{31}$ is the effective piezoelectric stress constant.

In addition, according to the constitutive equation and Gauss’ Law, the coupled electroelastic equation of the $i^{th}$ pair shown in figure~\ref{fig:shcema3} can be written as \cite{tol2016piezoelectric,erturk2011piezoelectric}:
\begin{equation}
\label{electroelastic equation}
\hskip0.7cm
C_{p}^{eq}\frac{dv_i(t)}{dt}+\frac{v_i(t)}{Z}+\chi\int_{x_i^L}^{x_i^R} \frac{\partial^3 w_i(x,t)}{\partial x^2 \partial t}  \,dx=0 
\end{equation}
where $C_{p}^{eq}$=11.5 nF is the equivalent capacitance of each pair of piezoelectric patches with the parallel connection.~$Z$ represents the external electric impedance. Superscripts R and L represent the right and left boundaries of the piezoelectric composite region.

Through a wave-based method by imposing the linear/angular displacement compatibility and force/moment equilibrium conditions at each interface, the complex wave amplitudes and the voltages can be simultaneously solved for each section in the waveguide~\cite{lin2021piezoelectric}. Then, the transmission ratios and phase shifts are obtained by dividing the propagating transmitted wave amplitude by the incident wave amplitude (see~\ref{derivation} for detailed derivation).

\subsection{Single-resonant shunt}
Our study starts by exploring the behavior of metasurface units with single-branch shunts containing only an inductive load $L$, resulting in an electric impedance $Z=\boldsymbol{i}\omega L$ for each pair.~Accordingly, an electromechanical resonance occurs in the vicinity of the electrical circuit resonance $\omega=(L C_{p}^{eq})^{-\frac{1}{2}}$, yielding adjustable control over dynamic responses due to the change in effective modulus~\cite{hagood1991damping}. 

\begin{figure}
\includegraphics[width=7cm]{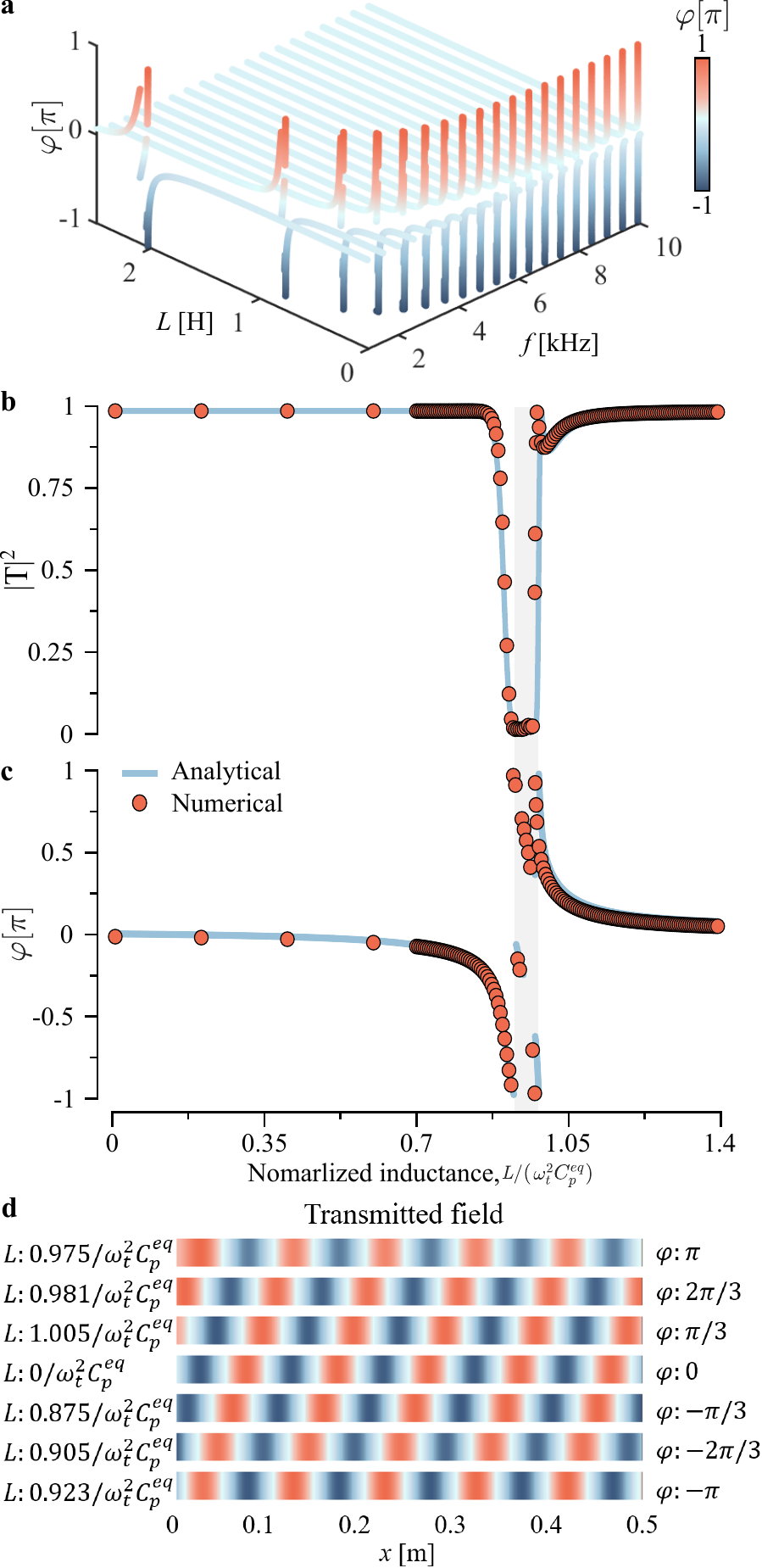}
\caption{Transmission~characteristics of the metasurface unit with single-resonant shunts.~(a) Frequency domain phase modulation capability under different inductive loads. Comparison between analytical and numerical (b) transmission ratios and (c) phase modulation results at 5 kHz, showing a good agreement even in the shaded region near the inductance generating resonance in the circuit. (d) Transmitted wavefields showing the phase variation from $-\pi$ to $\pi$ under different inductive loads.}
\label{fig:Phase_validate}
\end{figure}

We consider an incident flexural wave at target frequency $\omega_t$ propagates from the left-hand side of the waveguide along the positive $x$-direction, and explore the change in phase under different inductive loads, as demonstrated in figure~\ref{fig:Phase_validate}(a).~It is observed that we can introduce any desired phase shifts via the metasurface, which can be adaptable to operate in a wide low-frequency range (i.e., 1 to 10 kHz) by only tailoring the inductance loads.~Next, we further examine the transmission ratios and phase shifts at $f_t=\omega_t/(2\pi)=5$~kHz under different normalized inductance $L/(\omega_t^2 C_p^{eq})$.~The transmission ratio ($|T|^2$) drops around the inductance value corresponding to the resonant frequency in the circuit ($L/(\omega_t^2 C_p^{eq})$~=~1) as expected.~With a single resonant shunt, the phase modulation curve covers the $2\pi$ phase range without including the inductance range in the shaded region with low transmission ratios, as shown in figure~\ref{fig:Phase_validate}(b) and (c).~Hence, the metasurface unit can modulate the phase to desired values while maintaining high transmission ratios.~Furthermore, we numerically test the phase modulation ability of the metasurface unit under different inductive loads through simulations in COMSOL Multiphysics. A 2D model is developed to calculate the electromechanical response of the metasurface unit under an incident flexural wave at 5 kHz.~The piezoelectric effect is included in the Multiphysics module, and the perfectly matched layers (PMLs) are implemented at the ends to avoid the reflection.~A high-quality mesh is ensured by using mesh size of $\lambda/120$. The numerical results have an excellent agreement with the analytical results, as shown in figures~\ref{fig:Phase_validate}(b) and (c). Transmitted wavefields in figure~\ref{fig:Phase_validate}(d) clearly show the $2\pi$ phase coverage of the electroelastic metasurface unit with different normalized inductance, validating the accuracy of the established analytical model.~Therefore, the fully coupled wave-based electromechanical model can provide accurate phase information on the metasurface unit.

\subsection{Multi-resonant shunt}
The single-resonant piezoelectric shunt can only vary the phase velocity around the targeted single resonant frequency.~To endow the electroelastic metasurface with the capability to achieve simultaneous wavefront control over transmitted waves at distinct frequencies with the same electric impedance profile, we exploit multi-resonant piezoelectric shunts, as illustrated in figure~\ref{fig:shcema3}(b).
 It is pointed out in previous studies~\cite{sugino2017investigation,sugino2018design} that the bandgaps of electromechanical systems with piezoelectric shunts are characterized by the poles and zeros of the open-loop transfer functions~(OLTF) of the system based on the root locus method:
\begin{equation}
\label{eqn:oltf}
\hskip2cm
OLTF(s)=\frac{(1+\alpha)s+h(s)}{s^2(s+h(s))}
\end{equation}
where $\alpha$ represents the dimensionless electromechanical coupling coefficient and $h(s)=(Z C^p_{eq})^{-1}$ is the normalized circuit admittance. $Z=(\frac{1}{L_1 s}+\frac{1}{L_2 s+1/(C_2 s)})^{-1}$ is the electric impedance, and s is a complex variable in the Laplace domain. Here we tailor the position of locally resonant bandgaps in the system by tuning the poles' value. The characteristic equation of poles can be derived from equation~(\ref{eqn:oltf}), yielding:
\begin{equation}
\label{eqn:poles}
s^2(L_1C^p_{eq}L_2C_2s^4+(L_1C^p_{eq}+(L_1+L_2)C_2)s^2+1)=0
\end{equation}
There is a second-order pole at $s=0$ and two pairs of complex-conjugate poles at $s=\pm\boldsymbol{i}\omega_1$ and $s=\pm\boldsymbol{i}\omega_2$.~The locally resonant bandgaps occur around frequencies $\omega_{1,2}$, characterized by the non-trivial roots, where the transmission ratios and phase modulation profiles are sensitive to electric impedance change.~In this study, we design a multi-resonant shunted metasurface to achieve simultaneous control over the flexural waves at two distinct frequencies $f_1=\omega_{1}/(2\pi)=~5$~kHz, $f_{2}=\omega_{2}/(2\pi)=10$~kHz. We select $L_1=(\omega_m^2 C_p^{eq})^{-1}=39.2$~mH, $L_2=90.6$~mH and $C_2=6.29$~nF to guarantee the bandgaps emerge around target frequencies by solving equation~(\ref{eqn:poles}), where $\omega_{m}=(\omega_{1}+\omega_{2})/2$.

\begin{figure}
\includegraphics[width=6.5cm]{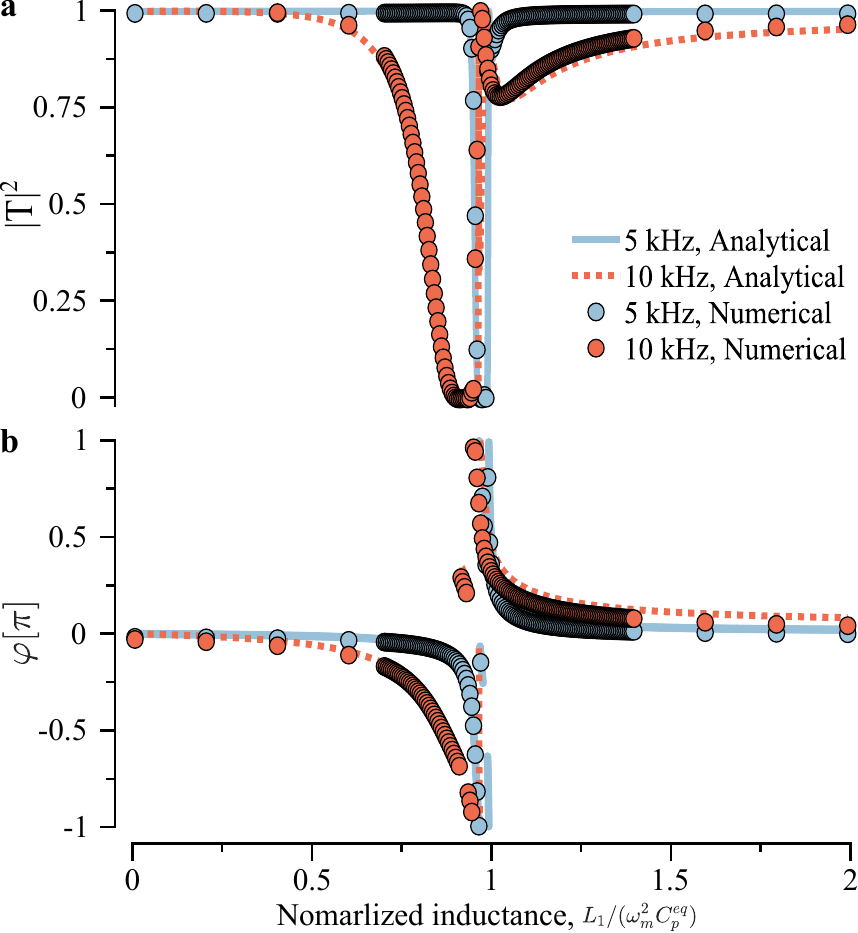}
\caption{Transmission characteristics of the metasurface unit with multi-resonant shunts. (a) Transmission ratios changing with normalized inductance, $L_1/(\omega_m^2 C_p^{eq})$. (b) Phase modulation curves of 5 kHz and 10 kHz flexural waves showing a similar trend.}
\label{fig:multiresonant_phase} 
\end{figure}

To explore the effect of electric impedance on the transmission ratios and phase modulation ability of the metasurface unit, we varied $L_1$ in the shunt circuit and fixed other parameters, as shown in figure~\ref{fig:multiresonant_phase}. It is observed in figure~\ref{fig:multiresonant_phase}(a) that transmission ratios of flexural waves at both target frequencies drop when normalized inductance,~i.e.,~$L_1/(\omega_m^2 C_p^{eq})$,~changes around 1, indicating the proposed arrangement can produce multiple resonant bandgaps around target frequencies.~In addition, the phase modulation curves in figure~\ref{fig:multiresonant_phase}(b) demonstrate that the multi-resonant shunt can introduce similar phase shift profiles for both waves by tailoring inductance, $L_1$, indicating the feasibility of achieving similar but not identical wavefront control over distinct frequencies in the desired range of interest.

\subsection{Experimental validation}
\begin{figure}
\includegraphics[width=8.2cm]{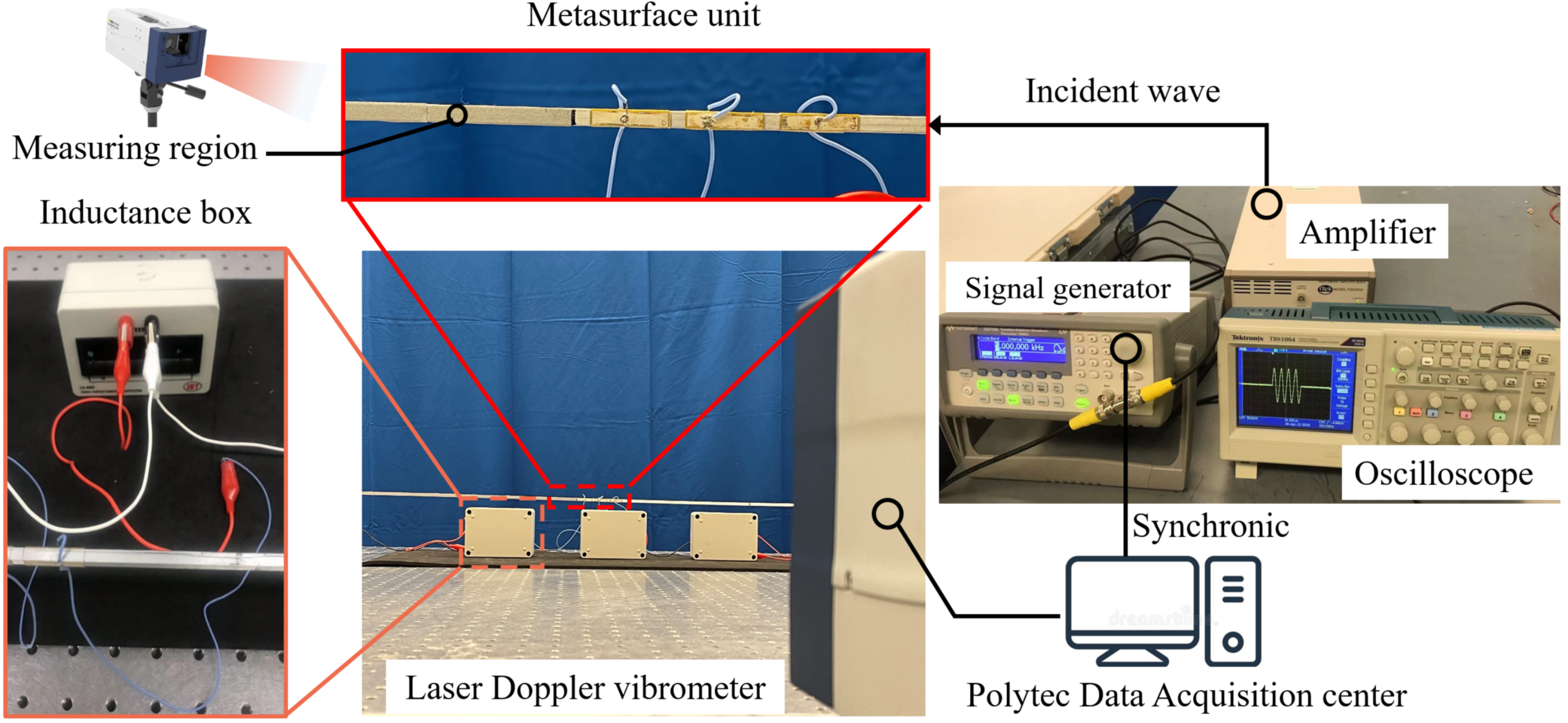}
\caption{Experimental set-up of the metasurface unit with single-resonant shunt.}
\label{fig:experimental_config}
\end{figure}

Next, we fabricate and experimentally test the phase modulation performance and transmission ratios of the metasurface unit. The experimental configuration with a single-resonant piezoelectric shunt is shown in figure~\ref{fig:experimental_config}.~Three piezoelectric patches (Steminc SM311) are equally distributed in the middle of a 6-foot-long aluminum beam to form the waveguide. The waveguide is sufficiently long to prevent the interference of reflected waves from the boundary.  The excitation signal is generated with the function generator (KEYSIGHT 33210A) and sent into the aluminum waveguide through a piezo driver (TREK PZD350A). A Polytec PSV-500 scanning laser Doppler vibrometer (LDV) is used to measure
the out-of-plane velocity response of the waveguide under different inductive shunts achieved via the inductance decade boxes (IET Labs, LS-400).

The mechanical and electrical losses in the system affect the phase modulation ability and transmission ratios of the metasurface unit. Therefore, we include the structural damping factor ($\eta_s=1\%$), the internal resistance of the inductance boxes ($R_l$), and the dielectric loss tangent ($\tan(\delta)$) in the model.~$R_l$ is directly measured via a multimeter.~The equivalent capacitance $C_{p}^{eq}$~=~11.5 nF and dielectric loss $\delta=10\%$ are identified from the energy harvesting performance of the piezoelectric patches.

\begin{figure}
\includegraphics[width=8cm]{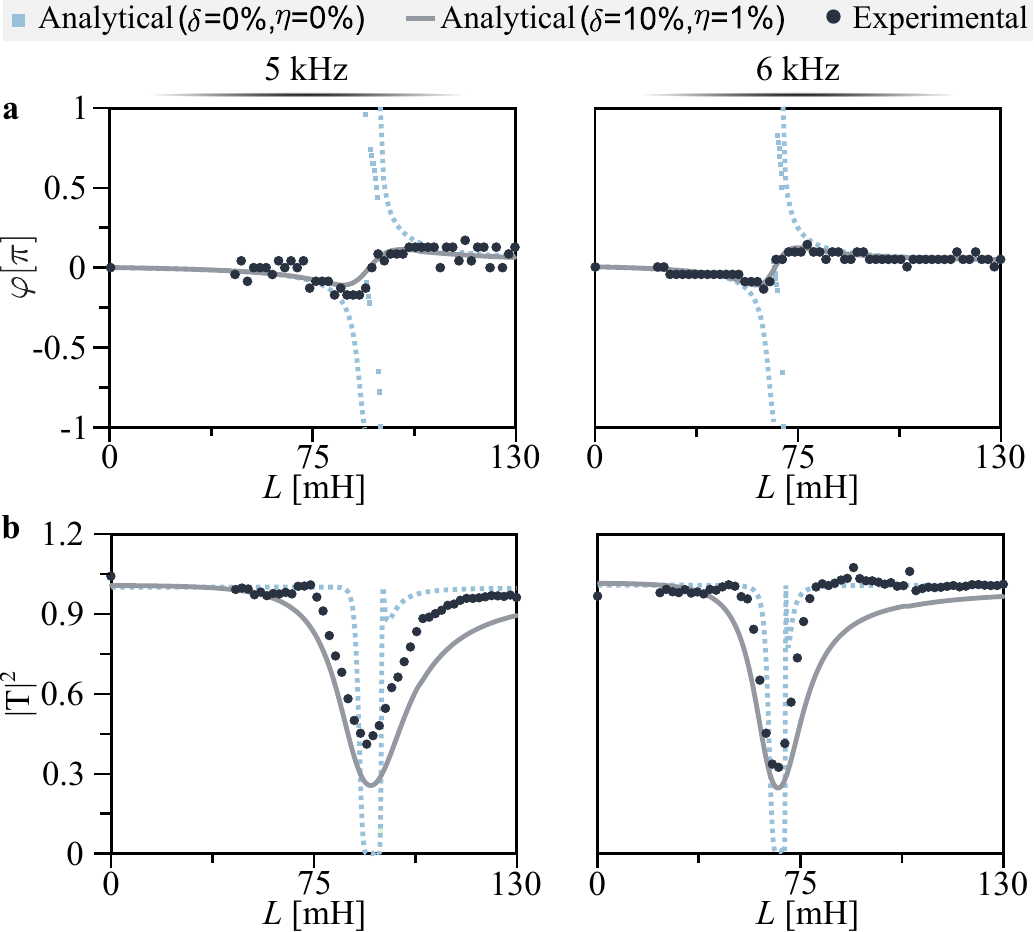}
\caption{Comparison~between~the~experimental~and~analytical~(a)~phase modulation results and (b) transmission ratios.}
\label{fig:phase_modulation_exp} 
\end{figure}

To test the tunable phase modulation performance of the metasurface unit with a single-resonant shunt, we excite low-frequency harmonic flexural waves (e.g., at 5 kHz and 6 kHz) propagating in the waveguide. The out-of-plane velocity response in the transmitted region is measured by LDV, as shown in figure~\ref{fig:experimental_config}.~Subsequently, the change in the peak's location of the velocity wavefields is traced to calculate the phase shift relative to the short circuit as plotted in figure~\ref{fig:phase_modulation_exp}(a).~The experimental phase modulation results and the analytical solutions match each other very well for both 5 kHz and 6 kHz cases.~It is observed that the phase change mainly occurs around the inductance $L=1/(\omega_t^2 C_p^{eq})$ mH, which are 88.1 mH for $f_t$ = 5 kHz and 61.2 mH for $f_t$ = 6 kHz. In addition, the phase modulation capability is reduced compared to an undamped system when the mechanical and electrical losses are included in the model.~Note that the phase modulation capability of the electroelastic metasurface can be improved through approaches that can reduce the electrical and mechanical losses in the system, e.g., using advanced piezo-bonding techniques or lossless inductance elements.

We also check the transmission ratios of the electroelastic metasurface.~In this case, four-cycle sine-burst flexural waves are sent into the waveguide.~The velocity of the incident and transmitted wave packets are first recorded by LDV and then post-processed with a fast Fourier transform(FFT) technique to obtain the transmission ratios, as plotted in figure~\ref{fig:phase_modulation_exp}(b). Compared to the lossless system, the transmission ratios increase, and the bandwidth becomes larger because the local resonators are imperfect due to the electrical and mechanical losses in the system. Overall, there is a good agreement between the experimental transmission ratios and the analytical prediction, validating the accuracy of the fully coupled analytical model.


\section{Electroelastic metasurface design and wavefront control}
Having the analytical model of the electromechanical waveguide validated through both numerical simulations and experiments, we design the electroelastic metasurface with ideal single-resonant and multi-resonant shunts to manipulate the wavefront of the transmitted waves in a tunable fashion.
\subsection{The metasurface with a single-resonant shunt for tunable wavefront tailoring}
We first design an electroelastic metasurface with single-resonant shunts to deflect the elastic wavefront of $\it{A_0}$ mode Lamb waves at different frequencies by only tuning the inductance. According to generalized Snell's law, to deflect a normally incident $\it{A_0}$ mode by an angle $\theta_t$ in the transmitted region, the phase profile along the metasurface should be a linear pattern satisfying:
\begin{equation}
\hskip2.5cm
\label{eqn:linear}
\varphi(y)=-y k_t\sin{\theta_t}+\varphi_c
\end{equation}
where $\varphi(y)$ represents the required phase value at location $\it{y}$ along the metasurface, and $\varphi_c$ is an arbitrary phase constant.

\begin{figure}
     \centering
     \includegraphics[width=7cm]{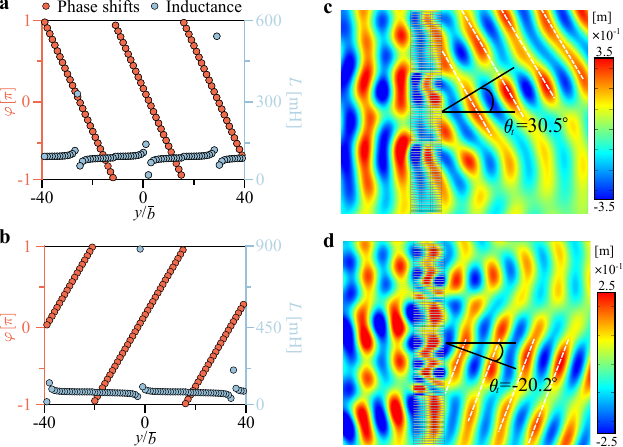} 
     \caption{Metasurface with single-resonant shunts for deflecting normally incident $A_0$ mode Lamb wave.~The phase profile and the required inductance values along metasurface for (a) 5 kHz and (b) 6 kHz wavefront deflection. $\bar{b}=7$ mm represents the distance between adjacent units.~The out-of-plane displacement wavefields of the wave-deflecting metasurface ((c):~5 kHz, (d):~6 kHz).}
     \label{fig:deflectingfinal}
\end{figure}

In order to demonstrate the tunability of the piezoelectric-based metasurface with a steering function, we only vary the inductive load to tailor the phase gradient such that the normally incident $A_0$ mode Lamb waves at 5 kHz and 6 kHz are deflected by 30 degrees and -20 degrees, respectively. 

The frequency-domain simulation is conducted to numerically test the metasurfaces with tailored phase gradients and corresponding inductance values, as shown in figure~\ref{fig:deflectingfinal}(a)-(d) for deflecting 5 kHz and 6 kHz waves, respectively. Figure~\ref{fig:deflectingfinal}(c) shows the out-of-plane displacement wavefield at 5 kHz, and the wavefront has been successfully deflected by an angle of $\theta_t=30.5^{\circ}$. Furthermore, the same metasurface can be adapted to deflect normally incident 6 kHz $A_0$ mode Lamb wave as shown in figure~\ref{fig:deflectingfinal}(d), where the wavefront is successfully deflected by an angle of $\theta_t=-20.2^{\circ}$.

Despite the small distortion in the wavefront due to the small variation in the transmission ratios along the metasurface and the limitation in aperture size, the electroelastic metasurface can achieve wave deflection of $A_0$ mode Lamb wave at different frequencies and at different angles by only tuning the inductance in the circuit as demonstrated with distinct case studies.

Aside from wave-deflecting metasurface, one of the most important metasurface designs in engineering practices is to achieve wave focusing, where high-intensity wave energy is desired towards various applications such as sensing and energy harvesting.

To localize wave energy at $(\it{f_x},\it{f_y})$ away from the metasurface, a semicircle equiphase surface is required in the transmitted region \cite{PhysRevLett.117.034302}, resulting in a hyperbolic-type phase profile as depicted in equation~(\ref{eqn:422}):
    
\begin{equation}
\label{eqn:422}
\hskip0.5cm
\varphi(y)=k_t (\sqrt{f_x^2+(y-f_y)^2}-\sqrt{f_x^2+f_y^2})+\varphi_c
\end{equation}

Similarly, $\varphi_c$ is an arbitrary phase constant added in the entire metasurface.~Note that adding/subtracting the phase constant $\varphi_c$ in the entire phase profile doesn't change the phase gradient along the metasurface, so wave energy still focuses at $(\it{f_x},\it{f_y})$.~On the other hand, $\varphi_c$ affects the required inductance profile of the metasurface and the transmission ratios of each metasurface unit. Hence, the wave intensity at $(\it{f_x},\it{f_y})$ varies with different $\varphi_c$ values, providing a new degree of freedom to optimize the focusing performance of metasurfaces.

Here, we evaluate the coverage of the wave-focusing metasurface at different radial focal angles. The Huygens–Fresnel principle is adopted to explore the optimal inductive loads. We approximate the response in the transmitted region as a superposition wavefield of an array of point loads adding at the end of each metasurface unit with the amplitude and phase profiles equal to $|T(y)|$ and $\varphi(y)$, respectively. The superposition wavefield can be given by equation~(\ref{eqn:444})~\cite{watanabe2014integral}:

\begin{equation}
\label{eqn:444}
\hskip0.3cm
\eqalign{w(x,y,t)= \sum_{i=1}^{79} \frac{-|T_i|e^{\boldsymbol{i} (\omega t+\varphi_i)}}{4\pi \sqrt{D\rho h_s} \omega} [&\frac{\pi \boldsymbol{i}}{2}H_0^{(2)}(k r_i(x,y)) \\
&+K_0(k r_i(x,y))]}
\end{equation}
where $\rho$ and $D=Y_sh_s^3/(12(1-\nu^2))$ are the density and the bending stiffness of the plate, respectively. $r_i(x,y)$ represents the distance between the point $(x,y)$ and the end of $i^{th}$ metasurface unit.~$H_0^{(2)}$ and $K_0$ are the Hankel function and modified Bessel function, respectively.~The optimal wave-focusing designs are obtained by sweeping $\varphi_c$ from 0 to 2$\pi$ with a preset optimization bound-constrained of the maximum inductance value less than 300 mH ($L\leq$ 300 mH).

\begin{figure}
\includegraphics[width=8cm]{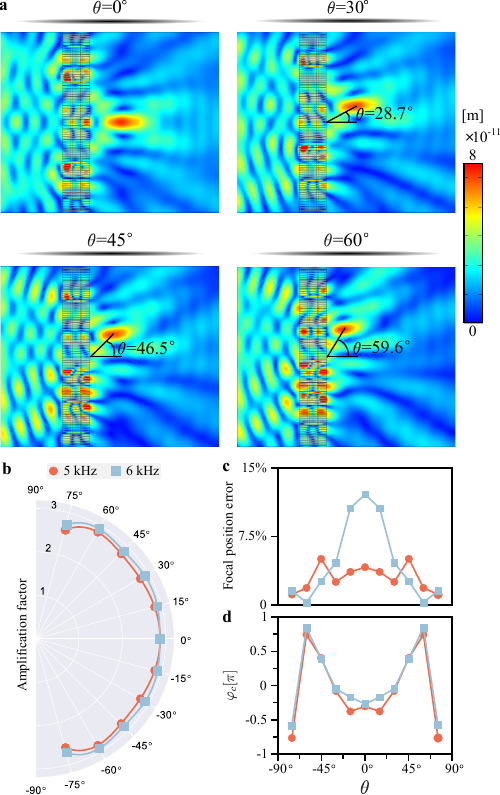}
\caption{Metasurface with single-resonant shunts for wave focusing.~(a) The out-of-plane displacement amplitude wavefields of the focusing metasurface design~of~5 kHz $A_0$ wave with different focal angles.~(b) The amplification factors of $|w|$ at different focal points $(f_x,f_y)=r(\cos{\theta},\sin{\theta})$. (c) The relative deviation of the numerical focal points away from the theoretical prediction. (d) The corresponding phase constant $\varphi_c$ for the optimal focusing performance.}
\label{fig:focusingfinal} 
\end{figure}

To demonstrate the versatility of the focusing metasurface, we tailor the phase gradient via varying the inductive load to localize normally incident $A_0$  mode Lamb wave at different focal points $(f_x,f_y)=r(\cos{\theta},\sin{\theta})$. Here, $\theta \in [-75^\circ,75^\circ]$ is the angle of focal points relative to the center of the metasurface, and $r$=101.6 mm is chosen around the wavelength of 5 kHz $A_0$ mode Lamb wave ($\sim$ 96.1 mm).

The frequency-domain simulation is conducted in COMSOL Multiphysics to numerically test the focusing metasurfaces for normally incident $A_0$ mode Lamb waves at different frequencies.~Here,~we first utilize it to localize the energy of normally incident 5 kHz $A_0$ mode at different angles of focal points.~Figure~\ref{fig:focusingfinal}(a) shows the out-of-plane displacement amplitude wavefields (i.e., $|w|$)  of the focusing metasurface for different angles of focal points, including $\theta=0^\circ$, $\theta=30^\circ$, $\theta=45^\circ$, $\theta=60^\circ$.~The focus can be clearly identified at the desired focal area for all cases. We quantitatively evaluate the focusing performance at different angles by calculating the amplification factor of the out-of-plane displacement amplitude relative to the baseline pure plate.~As shown in figure ~\ref{fig:focusingfinal}(b), the out-of-plane displacement of 5 kHz $A_0$ mode Lamb waves can be magnified nearly three times for any transmitted angles ranging from $-75^\circ$ to $75^\circ$. In addition, a small drop in amplification factor can be identified at a large angle, e.g., 75$^\circ$, due to the close distance between the focal points and the metasurface. 
 To evaluate the deviation of focal points from the theoretical prediction, we calculate the relative error of the focal position $\varepsilon=| \sqrt{f_{x,num}^2+f_{y,num}^2}-r |/r$ as plotted in figure~\ref{fig:focusingfinal}(c), which demonstrates a small deviation in focal position for all cases (i.e., $\varepsilon<6 \%$ for 5 kHz), where $(f_{x,num},f_{y,num})$ is the focal position obtained from numerical simulation results.  Also, the phase constants $\varphi_c$ corresponding to the optimal focusing designs are provided in figure~\ref{fig:focusingfinal}(d).

Likewise, we can tune the inductance to tailor the wavefront at 6 kHz to showcase its tunability. To this end, we utilize the metasurface to focus 6 kHz $A_{0}$ mode at different angles of focal points.~The out-of-plane displacement amplification factors are plotted in figure~\ref{fig:focusingfinal}(b), demonstrating a similar strong focusing capability as 5 kHz and achieving almost three times wave intensity for different transmitted angles.~The relative error of the focal position and optimal $\varphi_c$ are given in figure~\ref{fig:focusingfinal}(c) and figure~\ref{fig:focusingfinal}(d), respectively.~Our results show that the metasurface can maintain the high performance for different focusing requirements (e.g., focal positions and operating frequencies).

\subsection{The metasurface with a multi-resonant shunt for simultaneous tailoring of distinct wavefronts}
We further endow the electroelastic metasurface with the capability to achieve simultaneous wavefront control at distinct frequencies with the same electric impedance profile exploiting multi-resonant shunt. Based on the phase modulation curve of multi-resonant shunts in figure~\ref{fig:multiresonant_phase}(b), we utilize the proposed metasurface to achieve simultaneous wavefront control of normally incident flexural waves at 5 and 10 kHz. 

\begin{figure}
\includegraphics[width=8.2cm]{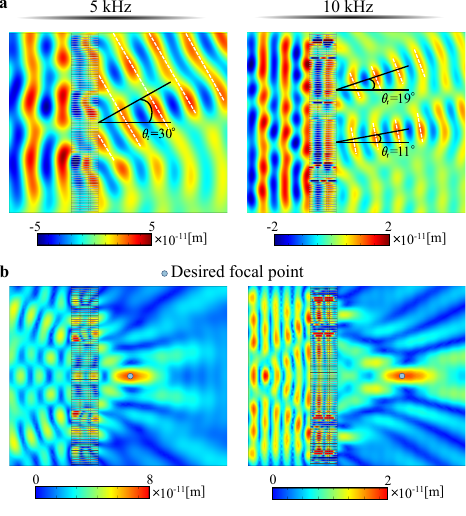}
\caption{Metasurface with multi-resonant shunts. (a) The out-of-plane displacement wavefields of the multi-resonant deflecting metasurface under different frequency incident waves demonstrating a smaller angle under higher frequency excitation.~(b) $|w|$ wavefields of the multi-resonant focusing metasurface under different frequency incident waves, validating that the wave intensity is localized at the desired region.}
\label{fig:multiresonant_simulation}
\end{figure}

The first type of multi-resonant metasurface is presented to simultaneously deflect distinct frequency waves in the transmitted region by different angles.~Here we tailor the metasurface to deflect a 5~kHz flexural wave by $\theta_{5\mathrm{k}}=30^{\circ}$ according to generalized Snell's law $\varphi(y)_{5\mathrm{k}}=-y k_{5\mathrm{k}}\sin\theta_{5\mathrm{k}}+\varphi_c$, where $\varphi_c=0.025\pi$ is selected to avoid large inductance value in the interpolation procedure. The same metasurface can simultaneously deflect a 10~kHz flexural wave by $\sin\theta_{10\mathrm{k}}=\sin\theta_{5\mathrm{k}}k_{5\mathrm{k}}/k_{10\mathrm{k}}=20.7^{\circ}$ due to similar phase gradient profiles of both frequencies indicated in the phase modulation curve figure~\ref{fig:multiresonant_phase}(b). Figure~\ref{fig:multiresonant_simulation}(a) shows the resulting displacement wavefields, where the 5 kHz flexural wave is successfully deflected by desired angle $\theta_{5\mathrm{k}}=30^{\circ}$, while two beams propagate in the transmitted region under 10~kHz excitation. One propagates along the direction angle of $\theta_{10\mathrm{k}}^1=19^{\circ}$ with only an 8$\%$ deviation from the theoretical prediction. The other is deflected by an even smaller angle of $\theta_{10\mathrm{k}}^2=11^{\circ}$ since the phase gradient of the 10~kHz wave changes slower relative to 5~kHz when the normalized inductance is less than 1, as shown in figure~\ref{fig:multiresonant_phase}(b), resulting in a smaller deflecting angle as we predicted by assuming identical phase modulation curve. Yet, it is successfully shown that the concept of multi-resonant metasurface can be used for simultaneous steering of waves propagating at different frequencies. The multi-resonant shunt electroelastic metasurface concept can be further adapted to other frequencies of interest by tailoring the electric loads.

In addition, we demonstrate that the proposed metasurface can be used to simultaneously localize the wave intensity of the propagating waves at both target frequencies.~Here the multi-resonant metasurface is designed to focus 5~kHz at $(f_x,f_y)_{5\mathrm{k}}=(101.6,0)$ mm based on equation~(\ref{eqn:422}).~The theoretical focal point under 10~kHz excitation is obtained by finding the optimal position $(f_x,0)_{10\mathrm{k}}$ to minimize the Euclidean norm of the phase gradient difference between two frequencies, $\parallel\varphi_{10\mathrm{k}}-\varphi_{5\mathrm{k}}\parallel_2$, yielding $(f_x,f_y)_{10\mathrm{k}}=(206.0,0)$~mm.~The appearance of a further focal point for higher frequency can be interpreted by rewriting equation~(\ref{eqn:422}) with the desired focal point $(f_x,0)$ into: $\varphi(y)=k_ty^2/(\sqrt{f_x^2+y^2}+f_x)+\varphi_c$. With similar phase gradient profiles, the higher frequency flexural wave with larger $k_t$ is always localized at a further position (larger $f_x$).~We examine the focusing metasurface performance through numerical simulations in COMSOL Multiphysics.~The $|w|$ wavefields under different frequency incident waves are shown in figure~\ref{fig:multiresonant_simulation}(b). The maximum wave intensities occur at 98.4 mm and 199.4 mm under 5 kHz and 10 kHz excitation, respectively, with only a $3\%$ deviation from the theoretical prediction. In addition, the maximum out-of-plane displacement is amplified by 2.8 times and 2.1 times, respectively, as compared to the baseline (i.e., pure plate).~The focusing performance under high-frequency excitation decreases due to the lower transmission ratios, as indicated in figure~\ref{fig:multiresonant_phase}(a), which could be enhanced by narrowing the bandgap range via tailoring the poles and zeros in the OLTF.

Note that all metasurface designs are realized by only changing the inductance in the circuit without any need for structural modification.~Therefore, the electroelastic metasurface can be easily tuned to satisfy different conditions and requirements.

\section{Conclusion}
In summary, we proposed and explored a piezoelectric-based elastic metasurface design with single-resonant and multi-resonant shunts to tailor the wavefront in the transmitted region. The electroelastic metasurface concept can be adapted to different frequencies of interest by tailoring the electric impedance.~Our innovation and findings include: (i) an almost equal wave-focusing effect with a nearly three times increase in the wave displacement for any designed angle of focal point ranging from $-75^\circ$ to $75^\circ$, (ii) a multi-resonance mechanism for achieving simultaneous wavefront control over flexural waves at distinct frequencies, (iii) a recipe for predicting the performance of metasurface with multi-resonant shunts.~With advanced theoretical, numerical, and experimental efforts, this study presents an unconventional wave control technique based on resonant-type electroelastic metasurfaces, offering new avenues to achieve tunable wavefront control without any structural modification.~Also, the multi-resonant shunted metasurface concept can be expanded to have simultaneous control over more than two target frequencies by adding additional resonant branches in the circuit, offering new potentials in designing tunable and adaptive devices for anomalous wavefront control.

\section*{Data Availability Statement}
The data that support the findings of this study are available upon reasonable request from the authors.

\section*{Acknowledgments}
This research was funded by the National Science Foundation under Grant No. CMMI-1933436.

\appendix

\section{Analytical formulation of the metasurface unit}
\label{derivation}
\subsection{Electromechanical model}
We establish a fully coupled electromechanical model and use the wave-based method to solve the wave propagation characteristics of the piezoelectric-based metasurface waveguide, as depicted in figure 1 (b). The 1-,2-,3-directions of piezoelectricity are coincident with the x-,y-,z- directions.

Since the substrate beam has a relatively small thickness compared to the wavelength (i.e., $\lambda/h_s>6$), the equation of motion of the flexural wave is governed by Euler-Bernoulli theory as depicted in equation~(\ref{eqn:11})~\cite{gonccalves2007numerical,fahy2007sound}.
\begin{equation}
\hskip2.2cm
\label{eqn:11}
    YI \frac{\partial^4 w}{\partial x^4} + m \frac{\partial^2 w}{\partial t^2} = 0
\end{equation}
where $w$ is the transverse displacement.~$YI$ and $m$ represent the bending stiffness and the mass per unit length, which are given below for both substrate beam and piezoelectric composite sections~\cite{erturk2011piezoelectric}:

\begin{equation}
YI=\cases{\frac{Y_{s}bh_{s}^{3}}{12} &\text{Substrate} \\
\frac{2b}{3}\left(Y_{s}\frac{h_{s}^{3}}{8}+c_{11}^{E}\left[-\frac{h_{s}^{3}}{8}+ \right.\right. \\ \left.\left. 
\hskip0.6cm
\left(h_{p}+\frac{h_{s}}{2}\right)^{3} \right]\right) &\text{Composite}\\}
\end{equation}

\begin{equation}
\hskip0.2cm
m=\cases{ \rho_{s}bh_{s} & \text{Substrate}  \\
\rho_{s}bh_{s}+2\rho_{p}bh_{p} & \text{Composite}\\}
\end{equation}
where $Y_s$ is Young's modulus of the substrate beam. $\rho$ and $c_{11}^{E}$ are the density and the elastic modulus of piezoceramic at the constant electric field, respectively. The subscripts and superscripts $p$ and $s$ stand for the corresponding parameter of the piezoceramic and substrate beam, respectively.

The bending moment in the composite section is equal to~\cite{erturk2011piezoelectric}:

\begin{equation}
\label{eqn:13}
\eqalign{M(x,t) =b(& \int_{-h_{p}-h_{s}/2}^{-h_{s}/2} T_{1}^{p}z\,dz+\cr
&\int_{-h_{s}/2}^{h_{s}/2} T_{1}^{s}z\,dz+\int_{h_{s}/2}^{h_{s}/2+h_{p}} T_{1}^{p}z\,dz)} 
\end{equation}
where $T_{1}$ represent the stress component in the $x$-direction. In addition, the reduced constitutive equations for a thin piezoelectric patch and substrate beam can be written as: 
\begin{equation}
\hskip2.5cm
\label{eqn:22}
T_{1}^{p}=c_{11}^{E}S_{1}^{p}-e_{31}E_{3}
\end{equation}
\begin{equation}
\hskip2.5cm
\label{eqn:23}
T_{1}^{s}=Y_{s}S_{1}^{s}
\end{equation}
where $e_{31}$ is the effective piezoelectric stress constant. $E_{3}$ is the instantaneous electric field in the z-direction. For the $i^{th}$ piezoelectric bimorph with parallel connection, the instantaneous electric fields are in the opposite directions in the top and bottom layers (i.e., $E_{3}=~-\frac{v_i}{h_{p}}$ in the top layer and $E_{3}=\frac{v_i}{h_{p}}$ in the bottom layer). $S_{1}$  represents the axial strain component and can be written as:
\begin{equation}
\hskip2.5cm
\label{strain}
S_1(x,z,t)=-z\frac{\partial^2 w}{\partial x^2}
\end{equation}

Substitute equation (\ref{eqn:22})-(\ref{strain}) into  equation (\ref{eqn:13}), the bending moment for the $i^{th}$ composite region can be simplified to equation~(\ref{coulped moment}).

In addition, according to the constitutive equation and Gauss’ Law, the coupled electroelastic equation of the $i^{th}$ bimorph shown in figure~\ref{fig:shcema3} can be written as equation~(\ref{electroelastic equation})~\cite{tol2016piezoelectric,erturk2011piezoelectric}.

\subsection{Wave-based solution of transmission characteristics}
\begin{figure}[h]
     \centering
     \includegraphics[width=0.45\textwidth]{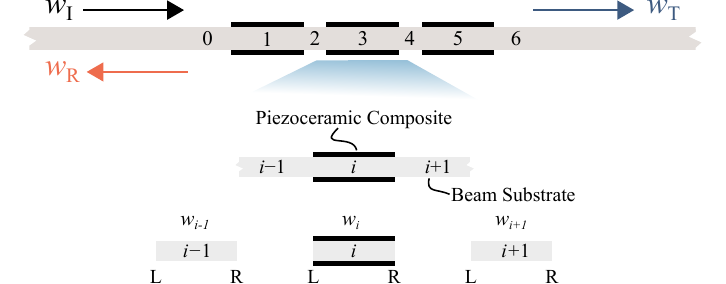}
     \caption{Schematic of the metasurface unit and piezoelectric bimorphs with superscripts R and L denoting right and left boundaries.}
     \label{3_patch}
\end{figure}

The metasurface unit is composed of six sections as shown in figure~\ref{3_patch}. For each section, the complete solution of equation~(\ref{eqn:11}) can be expressed as:
\begin{equation}
\hskip1cm
    \eqalign{w_i(x,t) = (&a_{1,i} e^{-\boldsymbol{i} k_i x} + a_{2,i} e^{-k_i x} + \cr &a_{3,i} e^{\boldsymbol{i} k_i x} + a_{4,i} e^{k_i x}) e^{\boldsymbol{i} \omega t}} 
    \label{eq:general_sol_disp}
\end{equation}
where $a_{1,i}, a_{2,i}, a_{3,i},$ and $a_{4,i}$ are the complex wave amplitudes and $i = 0,1,\dots,6$ denote corresponding composite and substrate sections along propagating direction. $k_i$ is the wavenumber defined as:
\begin{equation}
\hskip3cm
    k_i = \sqrt[4]{\frac{m \omega^2}{YI}}
\end{equation}

There are two propagating components and two evanescent components in equation~(\ref{eq:general_sol_disp}). The first (last) two terms in equation~(\ref{eq:general_sol_disp}) with a negative (positive) sign of the exponents denote the forward (backward) propagating and evanescent waves.
For an infinitely long waveguide, $a_{2,0} = 0$ is imposed in the incident region, and $a_{3,6} = a_{4,6} = 0$ are assumed in the transmitted region. 

Since the piezoelectric bimorph generates an additional moment term as given in equation~(\ref{coulped moment}) due to the inverse piezoelectric effect, the moment equilibrium conditions at the left and right boundaries of the piezoceramic composite region can be written as \cite{lin2021piezoelectric}:
\begin{equation}
\hskip0.8cm
    \label{eq:moment_eq_1}
\left(-YI\frac{\partial^2 w}{\partial x^2}\right)_{i-1}^{R}=\left(-YI\frac{\partial^2 w}{\partial x^2}\right)_{i}^{L}+\chi v_i
\end{equation}

\begin{equation}
\hskip0.8cm
     \label{eq:moment_eq_2}
\left(-YI\frac{\partial^2 w}{\partial x^2}\right)_{i+1}^{L}=\left(-YI\frac{\partial^2 w}{\partial x^2}\right)_{i}^{R}+\chi v_i
\end{equation}

In addition, the linear/angular displacement compatibility and force equilibrium conditions at each interface can be written as:
\begin{equation}
\hskip2.2cm
\boldsymbol{z}_{i-1}^{R}=\boldsymbol{z}_{i}^{L}, \boldsymbol{z}_{i+1}^{L}=\boldsymbol{z}_{i}^{R}
\end{equation}
where
\begin{equation}
\hskip1.2cm
\boldsymbol{z}=[w,\theta,Q]^{T}=[w,\frac{\partial w}{\partial x}, -YI\frac{\partial^3 w}{\partial x^3}]
\end{equation}

Accordingly, the complex wave amplitudes and the voltages can be simultaneously solved for each section in the waveguide by imposing the linear/angular displacement compatibility and force/moment equilibrium conditions at each interface together with the coupled electrical circuit equation as given in equation~(\ref{electroelastic equation}). And the transmission characteristics are obtained by dividing the propagating transmitted wave amplitude $a_{1,6}$ by the incident
wave amplitude $a_{1,0}$:
\begin{equation}
\hskip2.5cm
\label{eqn:transmission_coe}
    T = \frac{a_{1,6}}{a_{1,0}}=|T|e^{\boldsymbol{i}\varphi}
\end{equation}
where $|T|$ and $\varphi$ represent the transmission ratio and phase shift, respectively.

\bibliographystyle{dcu}
\bibliography{piezometa.bib}
\end{document}